\documentclass[12pt]{iopart}
  \usepackage{graphicx}

\begin{document}

\title[Small SQUID response]{The response of small SQUID pickup loops to magnetic fields}

\author{John R Kirtley,$^\dagger$ 
        Lisa Paulius,$^\ddagger$ 
        Aaron J Rosenberg,$^\dagger$ 
        Johanna C. Palmstrom,$^\dagger$ 
        Daniel Schiessl,$^\aleph$
        Colin L. Jermain,$^\star$ 
        Jonathan Gibbons,$^\star$ 
        Connor M. Holland,$^\dagger$ 
        Y.-K.-K. Fung,$^\ast$ 
        Martin E. Huber,$^\amalg$ 
        Mark B. Ketchen,$^\sharp$ 
        Daniel C. Ralph,$^\wedge$ 
        Gerald W. Gibson, Jr.,$^\ast$ 
        and Kathryn A. Moler$^\dagger$}
%
%\email{jkirtley@stanford.edu.}
\address{$\dagger$ Dept. of Applied Physics, Stanford University, Stanford, California 94305-4045}%\\This line break forced with \textbackslash\textbackslash

 \address{$\ddagger$ Physics Department, Western Michigan University, Kalamazoo, Michigan 49008-5252}%Lines break automatically or can be forced with \\

\address{$\aleph$ Attocube Systems AG,
K{\"o}niginstra{\ss}e 11a, 80539 Munich, Germany}

\address{$\star$ Dept. of Physics, Cornell University, Cornell, Ithaca, New York  14853}

\address{$\ast$ IBM Research Division, T.J. Watson Research Center, Yorktown Heights, New York 10598}

\address{$\amalg$ Dept. of Physics, University of Colorado Denver, Denver, Colorado 80217-3364}

\address{$\sharp$ OcteVue, Hadley, Massachusetts 01035}

\address{$\wedge$ Dept. of Physics, Cornell University, Cornell, Ithaca, New York  14853 and Kavli Institute at Cornell, Ithaca, New York, 14853}

\ead{jkirtley@stanford.edu}
\begin{abstract}
In the past, magnetic images acquired using scanning Superconducting Quantum Interference Device (SQUID) microscopy have been interpreted using simple models for the sensor point spread function. However, more complicated modeling is needed when the characteristic dimensions of the field sensitive areas in these sensors become comparable to the London penetration depth. In this paper we calculate the response of SQUIDs with deep sub-micron pickup loops to different sources of magnetic fields by solving coupled London's and Maxwell's equations using the full sensor geometry. Tests of these calculations using various field sources are in reasonable agreement with experiments. These calculations allow us to more accurately interpret sub-micron spatial resolution data obtained using scanning SQUID microscopy.
\end{abstract}

\maketitle

\section{Introduction}
The ultimate spatial resolution in SQUID microscopy has improved by about two orders of magnitude, from tens of microns to tenths of microns, over the last few decades\cite{rogers1983deo,vu1993dis,black1993mmu,kirtley1995hrs,veauvy2002smus,troeman2007nbn,finkler2010san,vasyukov2013scanning}. While this improvement in spatial resolution provides experimental opportunities, it also introduces new challenges for interpretation. For example, one is often interested in determining the absolute magnitude of the magnetic moments, electrical currents, or magnetic susceptibilities of the sample from the measured magnetic flux. This determination becomes increasingly difficult as the size of the sensor becomes comparable to the characteristic superconducting lengths. Modeling of SQUID response to magnetic fields has often relied on simplified models, such as assuming that the SQUID or the SQUID pickup loop captures all magnetic flux within a circle with an effective radius, sometimes adding an additional term for redirection of flux from the SQUID leads\cite{kirtley2012ssp}. Previous work has described effects that arise when the dimensions of the SQUID become comparable to the coherence length \cite{hasselbach2002msq}. In this paper we described a method for calculating a SQUID's response when its dimensions become comparable to the London penetration depth. This method solves coupled London's and Maxwell's equations, taking into account the full geometry of the field sensitive region. Although in this paper we apply and test this method  using large SQUIDs with small integrated pickup loops, the basic technique could also be readily used for small SQUIDs.

\section{Description of the calculation}
\label{sec:calculation}
Our calculations use a method developed by Brandt\cite{brandt2005tss}. We summarize this method here for completeness. The behavior of the three-dimensional super-current density ${\vec j}$ in a magnetic field ${\vec H}$ is described by London's second equation:
\begin{equation}
\nabla \times {\vec j} = -{\vec H}/\lambda^2,
\label{eq:3dlondon}
\end{equation}
where $\lambda$ is the London penetration depth. For a two-dimensional film of thickness $d$ in the $xy$ plane we integrate over $z$ to obtain
\begin{equation}
\nabla \times {\vec J} = -\vec{H}/\Lambda,
\label{eq:2dlondon}
\end{equation}
where ${\vec J}$ is the two-dimensional super-current density and $\Lambda \equiv \lambda^2/d$ is the Pearl length. Brandt defines a stream function $g(x,y)$ such that 
\begin{equation}
{\vec J}= -{\hat z} \times \nabla g = \hat{x} \frac{\partial g}{\partial y}-{\hat y}\frac{\partial g}{\partial x}.
\label{eq:stream}
\end{equation}
With this definition one can write London's second equation as
\begin{equation}
H_z(x,y) = \Lambda \nabla ^2 g(x,y)
\label{eq:hz}
\end{equation}
In addition, the stream function $g(x,y)$ can be thought of as a density of current dipoles. Then the total $z$-component of the field in the plane of a 2-d superconductor can be written as
\begin{equation}
H_z({\vec{r}}) = H_a({\vec{r}})+\int_S d^2{r'} Q({\vec r},{\vec{r}'})g(\vec{r}'),
\label{eq:hztotal}
\end{equation}
where $H_a({\vec{r}})$ is the externally applied field, and 
\begin{equation}
Q({\vec r},{\vec r}') = \lim_{z \rightarrow 0} \frac{2z^2-\rho^2}{4\pi(z^2+\rho^2)^{5/2}},
\label{eq:greens_function}
\end{equation}
with $\rho = |{\vec{r}-{\vec{r}'}}|$. Writing Eq. \ref{eq:hztotal} and \ref{eq:greens_function} as discrete sums:
\begin{equation}
H_z(r_i)=H_a(r_i)+\sum_j Q_{ij}w_jg(r_j),
\label{eq:discrete_hz}
\end{equation}
where $w_j$ is a weighting factor with the dimensions of an area, and
\begin{equation}
Q_{i\neq j} = \frac{-1}{4\pi |{\vec{r}}_i - {\vec{r}}_j|^3} \equiv -q_{ij}.
\label{eq:discrete_greens}
\end{equation}
$Q_{ij}$ is highly divergent for small values of $\rho$. Brandt handles this by noting that the total flux through the plane $z=0$ from any dipole source is zero in the absence of an externally applied field. Then for any ${\vec r}_i$ in the superconductor
\begin{equation}
0 = \int d^2 r' \, Q({\vec r}_i-{\vec r}') = \sum_j Q_{ij} w_j + \int _{\bar{S}} d^2 r' Q({\vec r}_i-{\vec r}').
\label{eq:zerosum}
\end{equation}
The discrete sum in Eq. \ref{eq:zerosum} is over the area inside the superconductor and the integral is over the area ($\bar{S}$) outside the superconductor. But the integral can be written as
\begin{equation}
\int_{\bar S} d^2 r' \, Q({\vec r}_i-{\vec r}') = \int_{\bar S} d^2 r' \frac{-1}{4 \pi | {\vec r}_i - {\vec r}' |^3} \equiv -C({\vec r}_i) = \oint \frac{d\phi}{4\pi R_i(\phi)},
\label{eq:circleint}
\end{equation}
where the last integral is over the angle $\phi$ between a fixed axis and a vector between the point ${\vec r}_i$ and a point on the periphery, and $R_i(\phi)$ is the length of this vector.  Returning to discrete sum notation,
\begin{equation}
Q_{ij} = (\delta_{ij}-1)q_{ij}+\delta_{ij}(\sum_{l \neq i}q_{il}w_l+C_i)/w_j
\label{eq:Qij}
\end{equation}
Eliminating $H_z$ from equation's \ref{eq:hz} and \ref{eq:discrete_hz} results in
\begin{equation}
H_a(r_i) = - \sum_j (Q_{ij} w_j- \Lambda \nabla^2_{ij})g(r_j)
\label{eq:ha}
\end{equation}
Inverting Eq. \ref{eq:ha} results in the solution for the stream function:
\begin{equation}
g({\vec r_i}) = -\sum_j K^{\Lambda}_{ij} H_a({\vec r_j}),
\label{eq:gfinal}
\end{equation}
with
\begin{equation}
K^{\Lambda}_{ij} = (Q_{ij} w_j - \Lambda \nabla ^2_{ij})^{-1}
\label{eq:Kfinal}
\end{equation}
The calculation strategy is then to first calculate the $K^{\Lambda}_{ij}$ matrix given the geometry and Pearl length $\Lambda$ from Eq.'s \ref{eq:circleint}, \ref{eq:Qij}, and \ref{eq:Kfinal} (in that order), calculate the stream function from Eq. \ref{eq:gfinal}, and then calculate the total field anywhere in the same plane for a given source field from Eq. \ref{eq:discrete_hz}. The field for any position with $z \neq 0$ is given by the discrete form of Eq.s \ref{eq:hztotal} and \ref{eq:greens_function}.

As pointed out by Brandt, it seems to work reasonably well to replace a detailed (and time consuming) calculation of $C_i$ from Eq. \ref{eq:circleint} with the analytical expression for a rectangular area $|x| \le a$, $|y| \le b$ which encloses the  superconducting shapes of interest:
\begin{equation}
C(x,y)=\frac{1}{4\pi} \sum_{p,q} [(a-px)^{-2}+(b-qy)^{-2}]^{1/2}
\label{eq:rectangle_C}
\end{equation}
with $p,q = \pm 1$.

We used Delaunay triangulation \cite{delaunay1934sphere} to tile our surfaces, and used a simplified version of a prescription by Bobenko and Springborn\cite{bobenko2006adl} to construct the Laplacian operator:
\begin{equation}
\nabla^2_{i,j} = \frac{1}{w} \sum_{j=1}^{N_i} (\delta_{i,j}-\delta_{i,i})
\label{eq:laplacian}
\end{equation}
where the sum is over the $N_i$  nearest neighbors of the $i^{th}$ vertex, and $w=ab/N_v$, with $ab$ the enclosing area (see Eq. \ref{eq:rectangle_C}) and $N_v$ the number of vertices in the triangulation. Eq. \ref{eq:laplacian} holds exactly for a square lattice, and seems to work well for a triangular lattice with sufficiently dense vertices.

Finally, Brandt provides a prescription for including externally applied currents. Assume for the moment that there is a delta function current $I$ at the inner edge of a superconducting shape with a hole in it. This is equivalent to applying an effective field
\begin{equation}
H^{\rm eff}_a = -I \sum_{\rm j\, in \, hole}(Q_{ij} w_j - \Lambda \nabla^2_{ij})
\label{eq:heff}
\end{equation}
The supercurrents generated in response to this field are described by the stream function
\begin{eqnarray}
g({\vec r}_i) &= - \sum_{\rm j \, in \, film} 
K^\Lambda_{ij} H^{\rm eff}_a ({\vec r}_j) \hspace{0.2in} &{\vec r_i }\, {\rm in \, film} \nonumber \\
&=I  & {\vec r_i }\, {\rm in \, hole} \\
&=0 & {\vec r_i }\, {\rm outside \, film} \nonumber
\label{eq:geff}
\end{eqnarray}
The fields generated by the current are then calculated from Eq. \ref{eq:discrete_hz} as before.

For multiple films we start from the film closest to the source, calculate its response to a source field, used the sum of the response field and the original source field as the source for the next closest film, etc. In principle one could also calculate the response of the full multiple-film structure self-consistently, but this procedure is very time consuming, converges slowly, and the results are close to the one we describe here for our geometry. Each film is taken to be two dimensional, with a $z$ position given by the average $z$ position of the film, but with the full geometry in the $xy$ plane.

We tested our program for several cases for which analytical expressions are available. For example, the self-inductance $L$ of a circular superconducting disk with inside diameter $d$ is known to approach $L=\mu_0 d$ as the outside diameter becomes large and the Pearl length $\Lambda \rightarrow 0$ \cite{ketchen1985dc}. Our calculation gives $L=4.95$ pH  for a circular disk with Pearl length $\Lambda=1$ nm, $R_{\rm in}$ = 2 $\mu$m and $R_{\rm out}$ = 50 $\mu$m. This is to be compared with $\mu_0 d$ = 5.02 pH. Similarly the mutual inductance between two co-planar, concentric, narrow circular wires can be calculated analytically. We calculate the mutual inductance between a larger ring with Pearl length 1 nm, inside radius 9.9 $\mu$m and outside radius 10 $\mu$m, and a smaller ring with inside radius 3.9 $\mu$m and outside radius 4 $\mu$m, to be 1.47$\times 10^3$ $\Phi_0$/A, where $\Phi_0 = h/2e$ is the superconducting flux quantum. This is to be compared with 1.40$\times 10^3$ $\Phi_0$/A calculated analytically using the geometric mean values $r_{\rm eff} = \sqrt{(r_{\rm in}^2+r_{\rm out}^2)/2}$ for the two ring radiuses. Our results are also in good agreement with those reported by Brandt \cite{brandt2005tss} in test cases with longer Pearl lengths.

\section{SQUID susceptometer pickup loop/field coil geometry}
\label{sec:geometry}
\begin{figure}
\includegraphics[width=6in, trim=0 0 0 0]{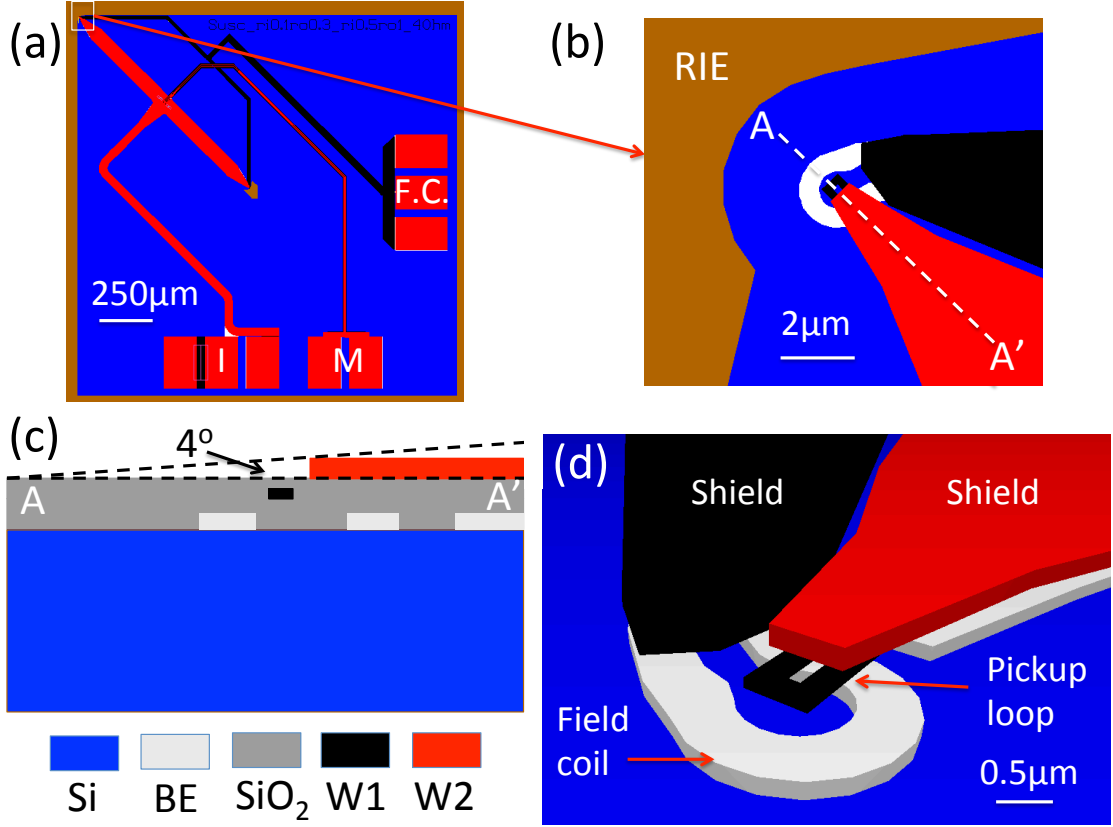}
\caption{Layout of the scanning SQUID susceptometers used to test our calculations. (a) Layout of the full chip. The bonding pads for the bias, modulation, and field coils are labelled $I$, $M$ and $F.C.$ respectively. (b) Pickup-loop/field coil region. A 10 $\mu$m deep etch region is labelled $RIE$. (c) Cross-section through the plane $A-A'$ indicated by the dashed line in (b). The region labelled $Si$ is the silicon substrate, $BE$ is the base electrode of the Nb-Al$_2$O$_3$-Nb trilayer, SiO$_2$ is the interlayer insulator, $W1$ is the first wiring level, and $W2$ is the second wiring level. (d) Three-dimensional view. The pickup loop, field coil, and shields are labelled.}
\label{fig:SUST_layouts}
\end{figure}

\begin{table*}[ht]

% title of Table
\centering 
% used for centering table
\begin{tabular}{| c | c | c |}
% centered columns (4 columns)
\hline
\hline                        %inserts double horizontal lines
Layer name & Material  &  Thickness (nm) \\
 [0.5ex]
% inserts table 
%heading
\hline  % inserts single horizontal line
BE & Nb & 160   \\
I1 & SiO$_2$ & 150   \\
W1  & Nb & 100  \\
I2  & SiO$_2$ & 130   \\
W2  & Nb & 200 \\
[1ex] % [1ex] adds vertical space
\hline %inserts single line
\end{tabular}
\label{table:layer_thicknesses} % is used to refer this table in
\caption{Layers composing the pickup loop/field coil region of our susceptometers. In order of deposition are the base electrode (BE), the first insulating layer (I1), the first wiring level (W1), the second insulating layer (I2), and the second wiring level (W2).}
\end{table*}

We test our calculations using results from our recently 
developed scanning SQUID susceptometers\cite{ketchen1989dfa,gardner2001ssq,huber2008gms} with sub-micron spatial resolution \cite{kirtley2016sss}. Briefly, these SQUIDs were fabricated using a planarized Nb-Al$_2$O$_3$-Nb trilayer process with two additional niobium wiring levels. Each SQUID had two resistively shunted junctions, with two pickup loops integrated in a gradiometric configuration through superconducting coaxial leads. Each pickup loop was surrounded by a 1-turn field coil. The layout of the full 2 mm $\times$ 2 mm chip is shown in Fig. \ref{fig:SUST_layouts}a. Since the bodies of our SQUIDs are well shielded by superconducting coaxes, only the region including the pickup loop and field coil is of relevance to the present paper. This area is displayed for our smallest pickup loops in Figure \ref{fig:SUST_layouts}(b). In this case the pickup loop was composed from the first wiring level (W1) as a square 180$^o$ bend with 0.2 micron linewidths and spacings. The pickup loop leads were shielded from below by the base electrode (BE) and above by the second wiring level (W2). Figure \ref{fig:SUST_layouts}(c) shows a cross-section through the layout along the dashed line labelled $A-A'$ in Fig. \ref{fig:SUST_layouts}(b). The 10 $\mu$m deep etch labelled $RIE$ in Fig. \ref{fig:SUST_layouts}(b) is 3 $\mu$m from the center of the pickup loop, and approximately 120 $\mu$m from the diced corner of the full chip. This means that without further processing the susceptometer must be aligned to an angle less than 5$^o$ for the etched edge to touch the sample first, and when this angle is less than 4$^o$ the top surface of the $W2$ pickup loop shield touches first. At all angles less than 4$^o$ the spacing between the top of the pickup loop and the surface of the sample is about 0.33 $\mu$m.

\section{Applications of the calculations to experiment}
\begin{figure}
\includegraphics[width=4in, trim=0 0 0 0]{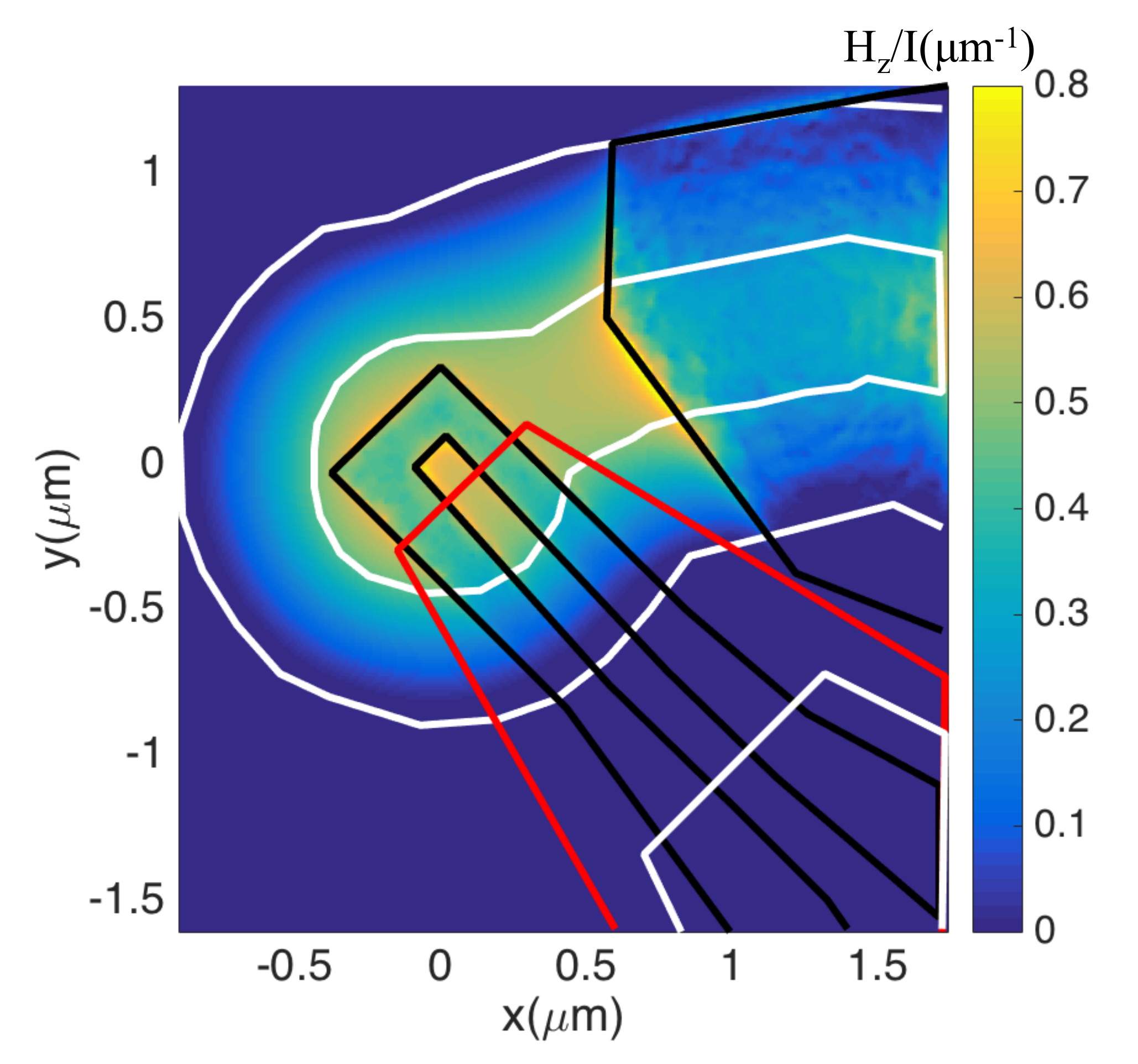}
\caption{Calculation of the pickup loop/ field coil mutual inductance. The colormap represents the calculated $z$-component of the magnetic field, divided by the current, in units of $\mu m^{-1}$. Overlaid on the image is the layout of the susceptometer, with the base electrode (BE) in white, the first wiring level (W1) in black, and the second wiring level (W2) in red. Integrating the field over the pickup loop area yields M=58.3 pH. The measured value was 69$\pm$7 pH.}
\label{fig:brandt_small_susceptometer}
\end{figure}

The resolution of a given SQUID susceptometer at a particular spacing from the field source depends on the type of source.  For example, the fields from a dipole fall off like $1/r^3$, those from a monopole source, such as a superconducting vortex, fall off like $1/r^2$, and those from a current source fall off like $1/r$. We have compared our calculations with experiments for all three types of field sources, as well as with measurements of the mutual inductance between the pickup loop and the field coil in our susceptometers, and susceptibility measurements of a superconducting pillar. All of the experimental data and calculations in this paper were for the 0.2 $\mu$m inside diameter pickup loop susceptometers with the pickup loop/field coil geometry displayed in Fig. \ref{fig:SUST_layouts}, assuming a London penetration depth for all superconducting films of $\lambda$=0.08 $\mu$m. The Pearl length $\Lambda$ for each film was calculated as $\Lambda=\lambda^2/d$, where $d$ was the film thickness. 

A source of systematic error in our measurements is the spacing between the surface of the susceptometer and the sample surface. We determined the $z$-piezo voltage required to keep the susceptometer in contact with the sample surface by feeding back on a deflection of the capacitance between the cantilever to which the susceptometer is attached and the sample mount.  Typically for a height series we took an initial scan in feedback mode to determine the contact $z$-position, and then backed off from contact by multiples of a fixed amount for successive scans, each scan taking 15-20 minutes. This procedure helped to reduce errors from piezo hysteresis and creep. The $z$-piezo displacement was calibrated by imaging a step of known height in topography in feedback mode. We estimate our uncertainties in the $z$-position to be approximately 0.1 $\mu$m.

\subsection{Field coil/pickup loop mutual inductance}
\label{sec:mutual}
One test of the quality of a susceptometer is to measure the mutual inductances between the field coil/pickup loop pairs. For example, if there is a short between the first and second wiring level in the coaxes leading to the pickup loops, the susceptometer critical current will modulate as expected using the modulation coil, but will have very small mutual inductance between one or both of the field coils and the SQUID. In addition, this mutual inductance can be used to normalize estimates of, for example, the London penetration depth or Pearl length of a superconducting sample from scanning SQUID susceptibility measurements\cite{kirtley2012ssp}. Figure \ref{fig:brandt_small_susceptometer} displays the z-component of the magnetic fields generated by current through the field coil, at the level of the pickup loop, calculated using the procedure outlined in Sec. \ref{sec:calculation}, using the geometry described in Sec. \ref{sec:geometry}. The mutual inductance between the field coil and the pickup loop, obtained by integrating the flux through the geometric centers of the pickup loop, is calculated to be $M=58.3$ pH. This is about 15\% lower than the measured value of 69$\pm$7 pH for these devices. It is not clear what the source of this discrepancy is. We speculate that the effective areas of the pickup loops are slightly (15\%, or 7\% in linear dimensions) larger than laid out because of the etching process, or that it is more appropriate to integrate over a larger region than given by the geometric centers of the pickup loop to determine the flux through the SQUID. Such an underestimate of the pickup loop area would lead to fit values for the spacing between sensor and sample that are too small, as appears to be the case for the vortex measurements in Section \ref{sec:vortex}.

\subsection{Dipole source}
\label{sec:dipole}

\begin{figure}
\includegraphics[width=6in, trim=0 0 0 0]{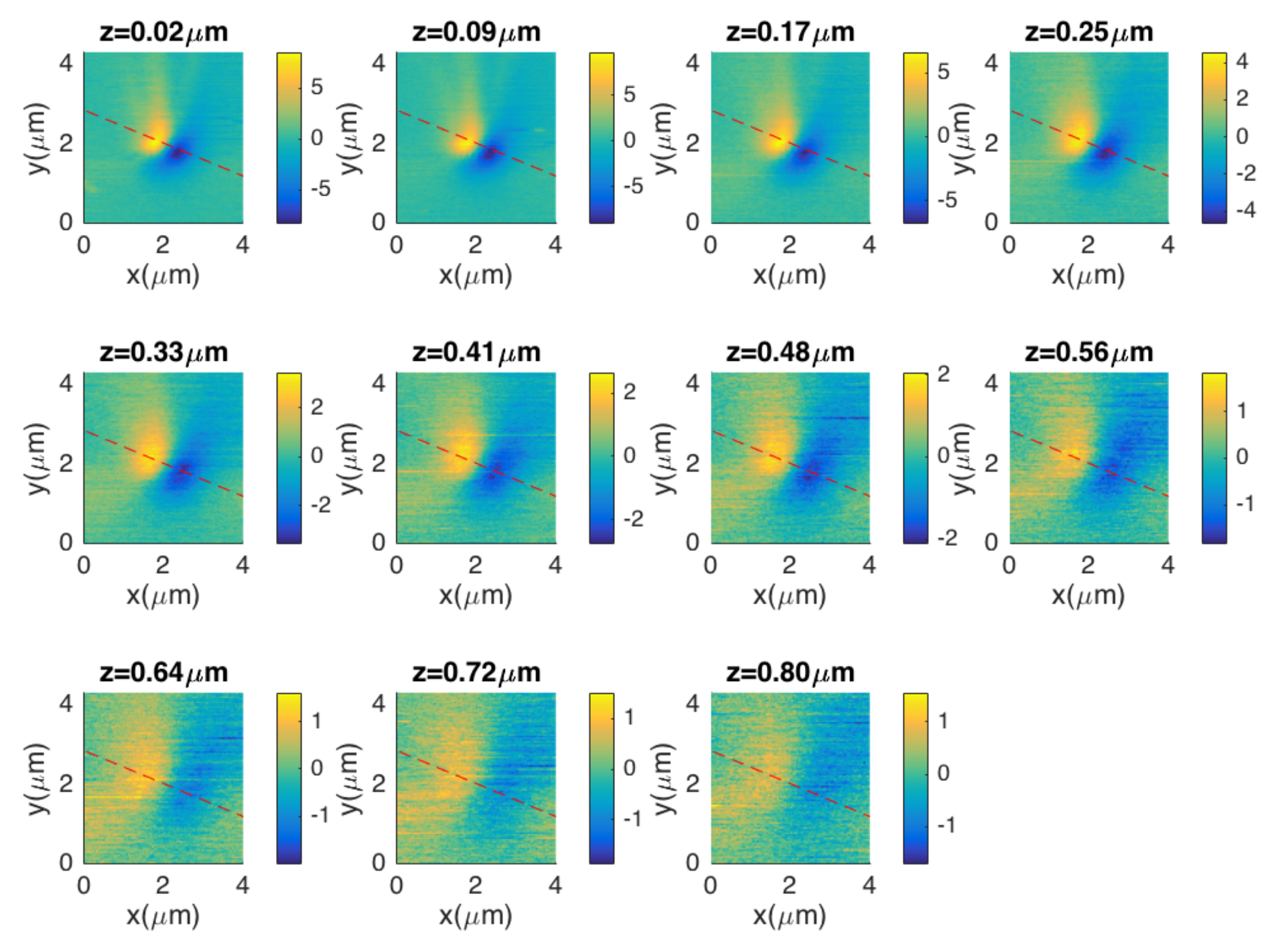}
\caption{SQUID magnetometry images of a magnetic nanoparticle  as a function of spacing between the susceptometer and sample surfaces. Each image is of an area 2 $\mu$m$\times$ 2 $\mu$m. The colormap scales are in units of $10^{-4}\Phi_0$.}
\label{fig:dipole_images}
\end{figure}

Figure \ref{fig:dipole_images} displays scanning SQUID magnetometry images from a CoFeB(4)/Pt(4) magnetic nanoparticle, where the numbers in parentheses are thicknesses in nm. We believe this nanoparticle was smaller than the $\approx$0.5 $\mu$m resolution of our susceptometer.
%The particle was nominally an ellipse with a semi-major axis $a=$25 nm, semi-minor axis $b$= 6.25 nm, and saturation magnetization of $M_s$=2.2$\pm$0.1$\times 10^6$ A/m, yielding a saturation moment of $\pi a b t M$=4.7$\times 10^5 \mu_B$, with t=4 nm the thickness of the magnetic layer.

\begin{figure}
\includegraphics[width=6in, trim=0 0 0 0]{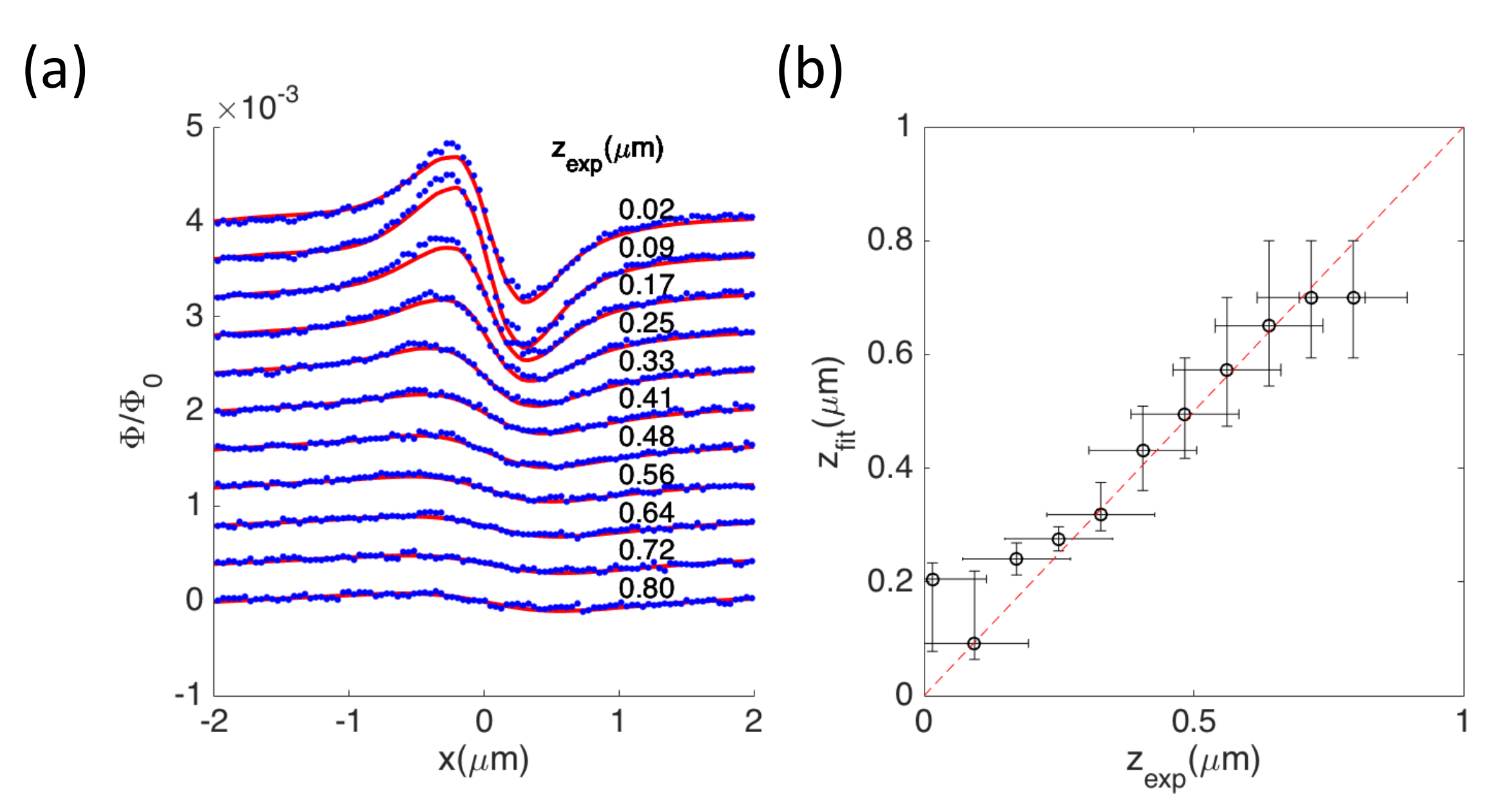}
\caption{Fits of the calculations of Sec. \ref{sec:calculation} to cross-sections along the dashed lines of Fig. \ref{fig:dipole_images}, using the total nanoparticle moment $|m|$ and the susceptometer surface to sample surface spacing $z_{\rm exp}$ as fitting parameters. (a) The dots are cross sections through the data, and the solid lines are fits assuming an in-plane dipole moment oriented at -15$^o$ to the horizontal axis, with a global fit moment $|m|$=1.7$\times 10^6\mu_B$, with the experimental spacings $z_{\rm exp}$ as labeled. Each successive curve is offset vertically by 4$\times 10^{-4} \Phi_0$ for clarity. (b) Plots  of the fit values $z_{\rm fit}$ vs. spacings $z_{\rm exp}$. The dashed line represents $z_{\rm fit}=z_{\rm exp}$.}
\label{fig:dipole_cross_section}
\end{figure}

The dots in Figure \ref{fig:dipole_cross_section} display
cross-sections through the dashed lines in the magnetometry images of Fig. \ref{fig:dipole_images}. The solid lines are fits to the data, assuming the nanoparticle is a point dipole source oriented in-plane, using the angle relative to the horizontal axis (15$^o$), total moment $|m|$ and the sample-susceptometer spacing $z_{\rm fit}$ as fitting parameters. The moment $|m|$ which gives the best global fit is $|m|=1.7\pm 0.6 \times 10^6 \mu_B$. %about 3.5 times larger than the calculated moment. We believe that in fact this particle was larger than intended, although still small enough to be resolution limited.

Figure \ref{fig:dipole_cross_section}(b) displays the best fit values for $z_{\rm fit}$, keeping the moment fixed at $|m|=1.7\times 10^6 \mu_B$, as a function of the experimental spacing $z_{\rm exp}$. These fits are consistent with the  dashed line $z_{\rm fit}=z_{\rm exp}$, which would be expected if the misalignment angle between the SQUID and sample was such that, as intended, the top surface of $W2$ directly above the pickup loop was in contact with the nanoparticle at $z_{\rm exp}=0$.

\subsection{Monopole source}
\label{sec:vortex}
\begin{figure}
\includegraphics[width=6in, trim=0 0 0 0]{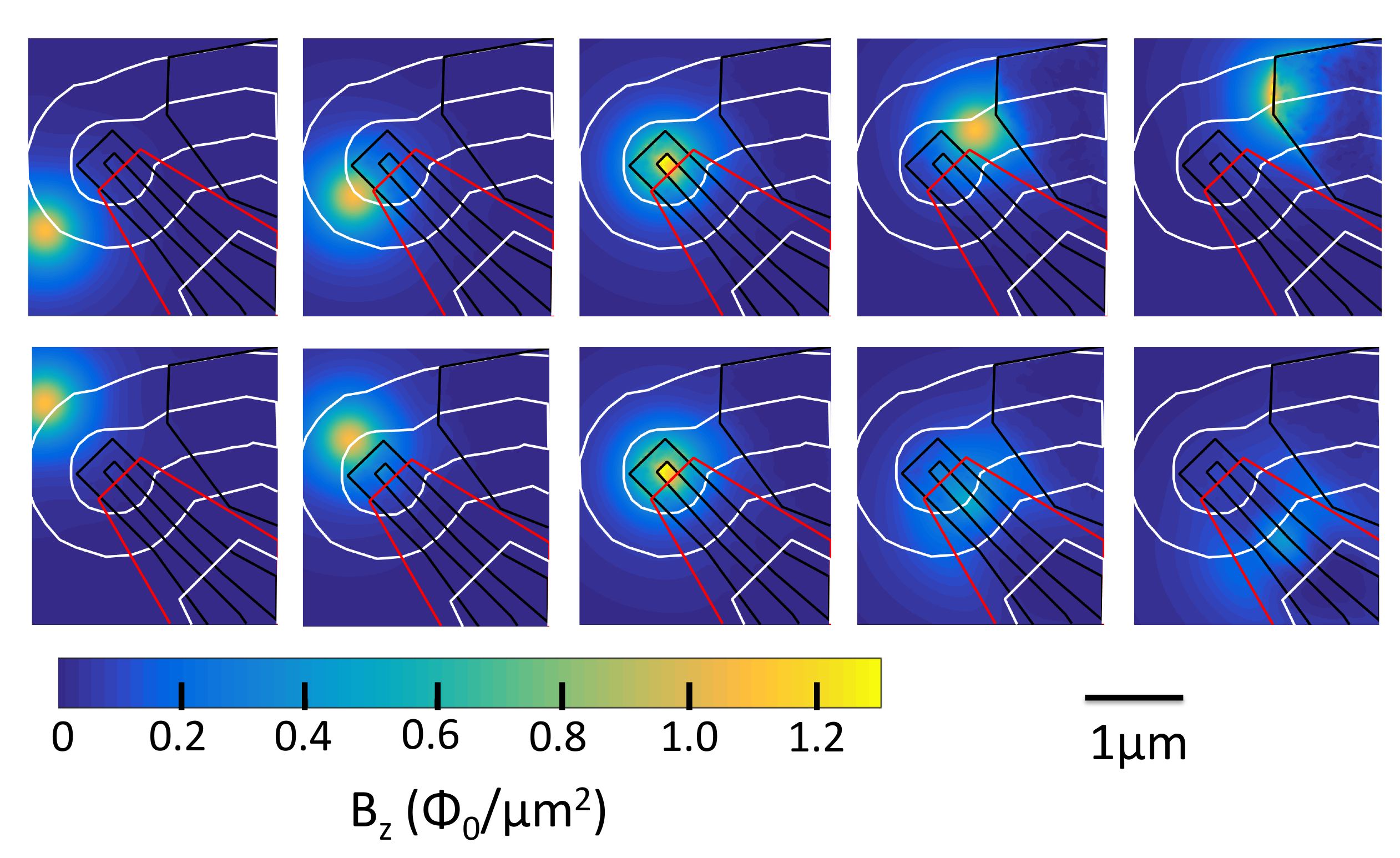}
\caption{Calculated response of our smallest pickup loop susceptometer to a superconducting vortex, with the upper surface of the sensor ($W2$) in contact with the sample surface. The total magnetic fields are calculated at the level of the pickup loop. The vortex center was spaced by -1 $\mu$m, -0.5 $\mu$m, 0 $\mu$m, 0.5 $\mu$m and 1 $\mu$m from the center of the pickup loop along a line perpendicular (top row) or parallel (bottom row) to the pickup loop leads.}
\label{fig:Brandt_vortex_images}
\end{figure}

\begin{figure}
\includegraphics[width=6in, trim=0 0 0 0]{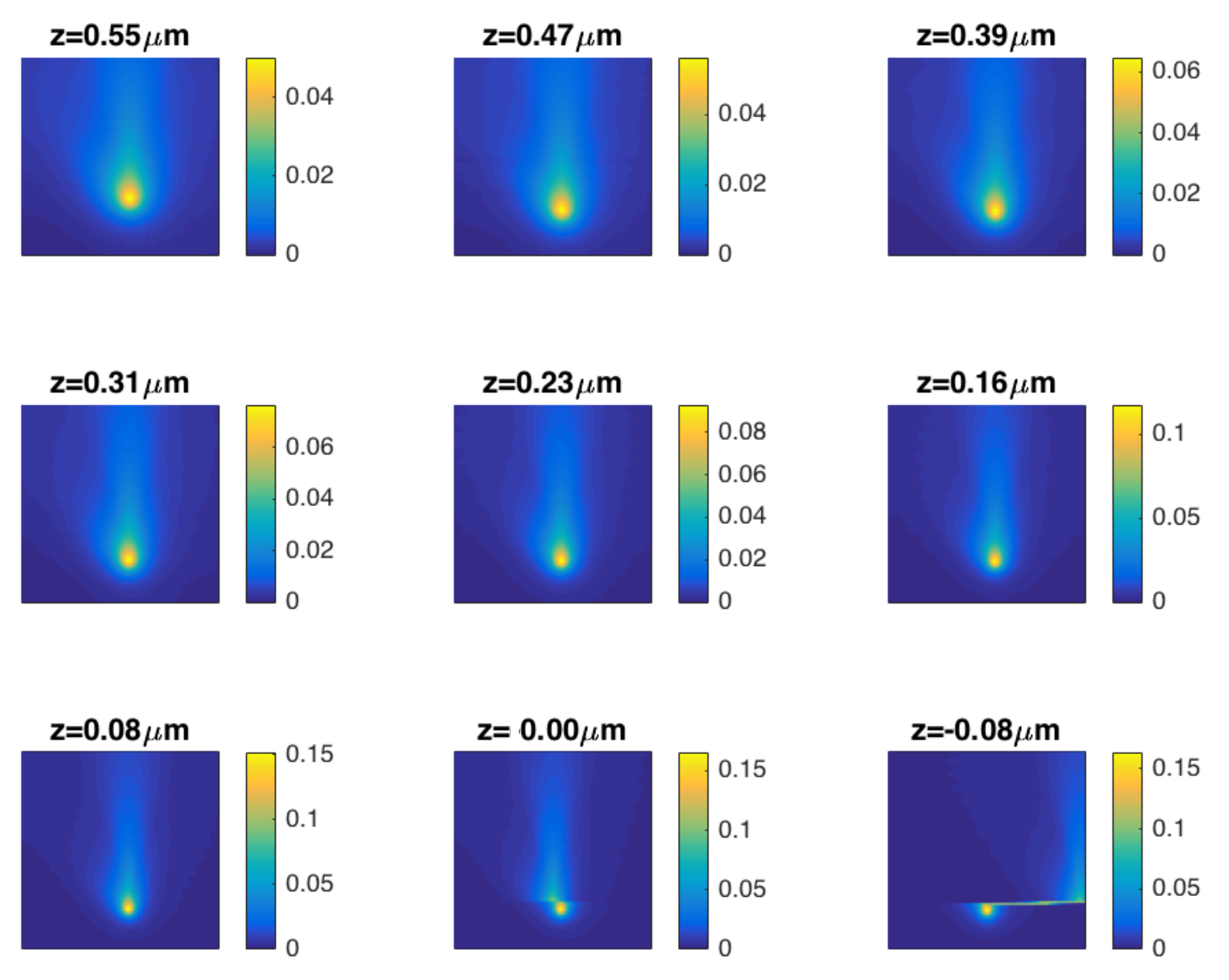}
\caption{Scanning SQUID magnetometry images of a superconducting vortex trapped in a 0.4 $\mu$m thick Nb film, for various spacings between the sample and susceptometer. Each image is of an area 4 $\mu$m by 4 $\mu$m. The colormap scales are in units of the superconducting flux quantum $\Phi_0=h/2e$. When the sensor comes into contact with the Nb surface the vortex moves in the slow scan (horizontal) direction. The pickup loop leads were oriented vertically. The negative number label $z$=-0.08 microns in the lower right image implies that the cantilever to which the SQUID is attached is bending.}
\label{fig:vortex_images}
\end{figure}

Another test of our modeling is the imaging of superconducting vortices. Figure \ref{fig:Brandt_vortex_images} displays calculations of the fields at the level of the pickup loop from a point source  magnetic monopole with total flux $\Phi=\Phi_0=h/2e$, the superconducting flux quantum, for scans perpendicular (upper row) and parallel (lower row) to the leads. One can see from these images that the vortex fields are ``focused'' by Meissner screening when  the pickup loop center is directly above the vortex, and that the vortex fields are redirected away from the field coil by the $W1$ shield, and from the pickup loop by the $W2$ shield.
 Figure \ref{fig:vortex_images} displays magnetometry images of a superconducting vortex trapped in a 0.4 $\mu$m thick Nb film, imaged at 4.2K at a series of spacings between the surface of the susceptometer and the surface of the sample. As this spacing decreases, the vortex image becomes more intense and more sharply defined, until when the SQUID tip is in direct contact with the sample surface the vortex is pushed in the slow scan (horizontal) direction \cite{kremen2016mci}. 

\begin{figure}
\includegraphics[width=6in, trim=0 0 0 0]{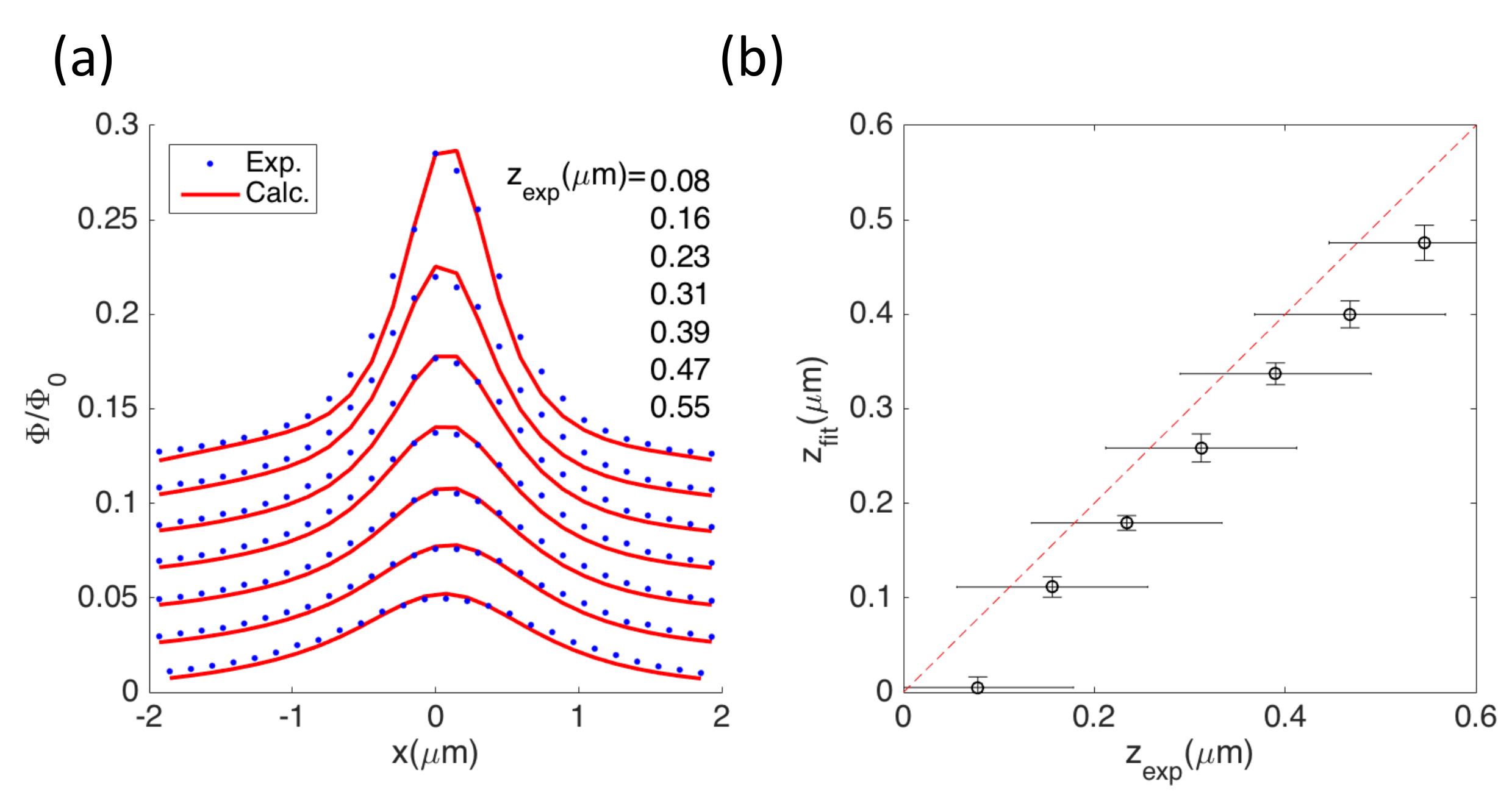}
\caption{(a) The dots are cross-sections in the horizontal direction (perpendicular to the pickup loop leads) of the vortex magnetometry images of Fig. \ref{fig:vortex_images}. The lines are fits to the calculations of Sec. \ref{sec:calculation} with the spacing $z_{\rm fit}$ between the upper surface of the sensor and the sample surface  as a fitting parameter. (b) Plots of the fit spacing $z_{\rm fit}$ vs. the experimental spacing $z_{\rm exp}$. The dashed line represents $z_{\rm exp}=z_{\rm fit}$.}
\label{fig:vortex_cross_section}
\end{figure}

The dots in Figure \ref{fig:vortex_cross_section}(a) display cross-sections in the horizontal direction through the vortex images of Fig. \ref{fig:vortex_images}. Successive curves are offset by 0.02 $\Phi_0$ for clarity. The solid lines in Fig. \ref{fig:vortex_cross_section}(a) are fits to this data assuming the vortex is a point monopole source with total flux $\Phi_0$, with the spacing between the sample and susceptometer surfaces $z_{\rm fit}$ as a fitting parameter. Figure \ref{fig:vortex_cross_section}(b) displays the best fits for $z_{\rm fit}$ plotted vs. the experimental spacing $z_{\rm exp}$, assuming the SQUID is in contact with the sample just when the vortex starts moving. As can be seen in Fig. \ref{fig:vortex_cross_section}(b) the fit values $z_{\rm fit}$ are slightly {\it below} the dashed line $z_{\rm fit}=z_{\rm exp}$. This is surprising, since the fields above the surface of a superconductor of a vortex are well approximated by a point monopole source a distance $\lambda$ {\it below} the surface\cite{pearl1966ssv}. This implies that $z_{\rm fit}$ should be larger than $z_{\exp}$ by about $\lambda=0.08 \mu m$ for niobium. We believe that this discrepancy may be the result of our modeling using too small a value for the effective area of the pickup loop, as discussed in Sec. \ref{sec:mutual}.

\subsection{Current source}
\label{sec:currents}
\begin{figure}
\includegraphics[width=6in, trim=0 0 0 0]{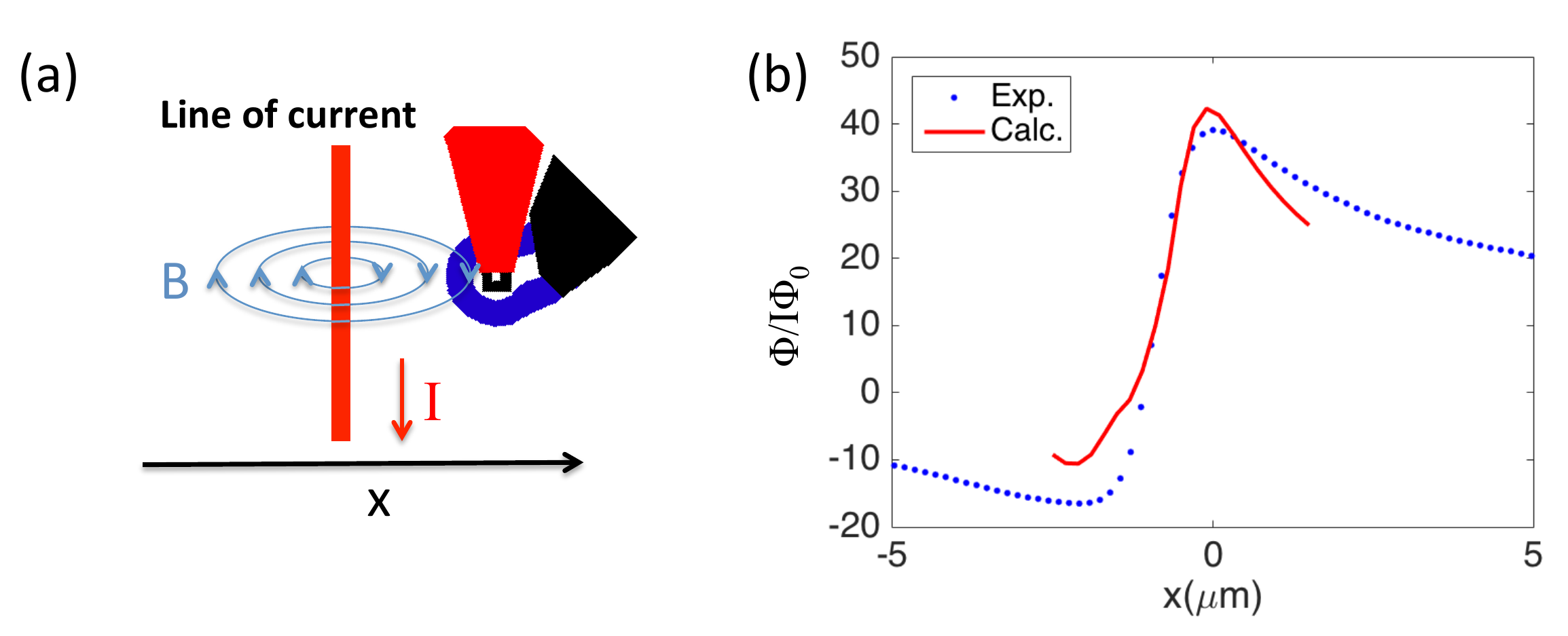}
\caption{(a) Schematic of current induced magnetometry imaging of currents in a thin wire. (b) The solid line is a current induced magnetometry cross-section, with the SQUID sensor in contact with the sample,  of a 0.8 $\mu$m wide current carrying strip. The open circles are the calculations of Sec. \ref{sec:calculation}, assuming the sensor is in contact with the sample surface.}
\label{fig:current_data}
\end{figure}

Figure \ref{fig:current_data}(a) illustrates the geometry we used for testing the response of our susceptometer to a line of current. The sample was a thin strip of Pt 0.8 $\mu$m wide, carrying a sinusoidally varying current with amplitude 1 mA at 500 Hz. The pickup loop leads were parallel to the current carrying strip, and perpendicular to the scan direction. The measurements were taken with the susceptometer in contact with the sample in feedback mode. The solid line in Fig. \ref{fig:current_data}(b) plots the SQUID response. The open symbols are the calculations as described in Sec. \ref{sec:calculation}, assuming the susceptometer was in contact with the sample, with no adjustable parameters. Although the agreement between experiment and calculation is reasonably good for the magnitude of the response, the cross-section is smeared relative to experiment, perhaps because of the finite width of the Pt strip.

\subsection{Susceptibility}
\label{sec:susceptibility}
\begin{figure}
\includegraphics[width=6in, trim=0 0 0 0]{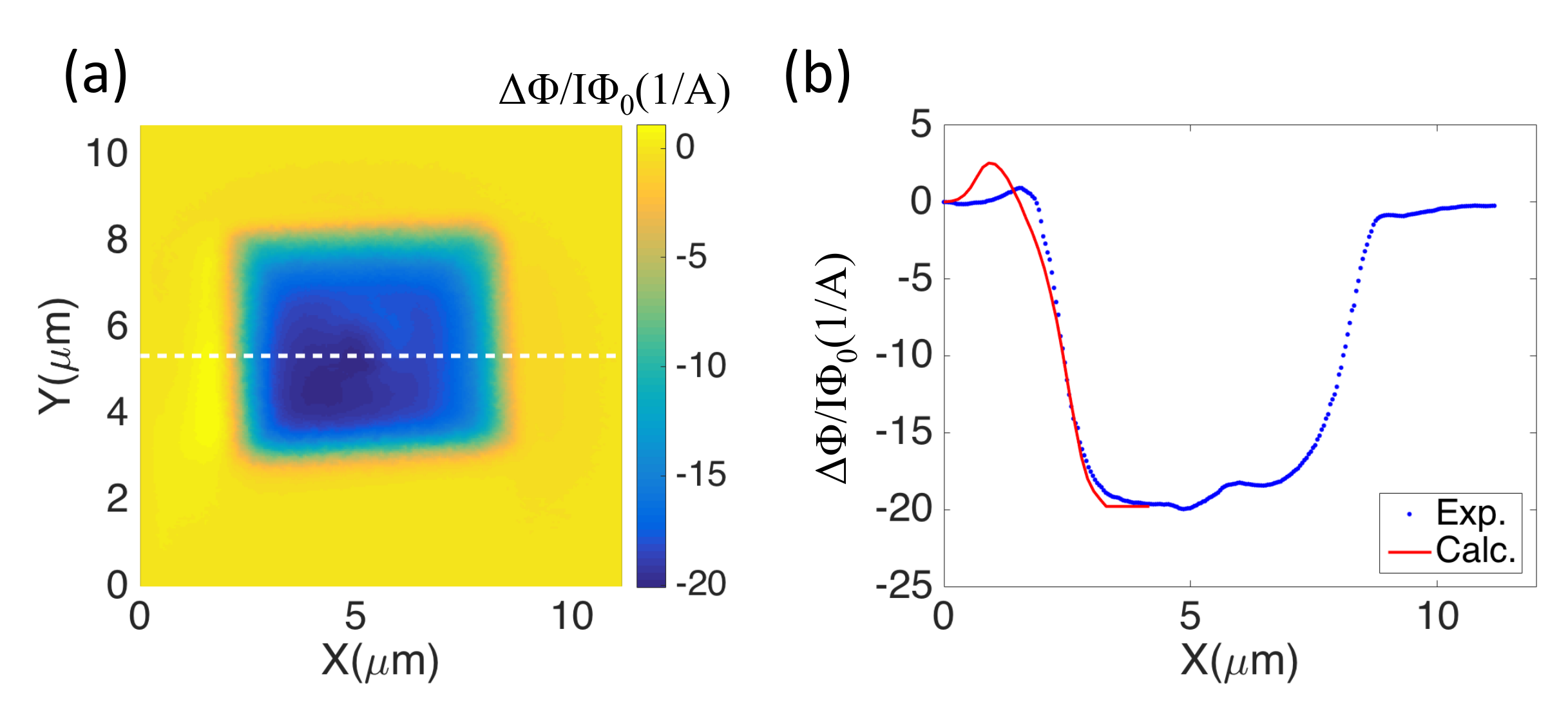}
\caption{(a) Susceptibility image of a 6 $\mu$m wide square niobium pillar. The colormap scale is in units of $\Phi_0/A$. (b) The solid line is a cross-section along the dashed line in Fig. \ref{fig:dot_susc_data}(a). The open circles are the calculations of Sec.\ref{sec:calculation}, with a fit value for the spacing between the susceptometer and sample surfaces of 0.8 $\mu$m.}
\label{fig:dot_susc_data}
\end{figure}

Finally, Figure \ref{fig:dot_susc_data}(a) shows a susceptibility image of a 6 $\mu$m square pillar, composed of a stack of alternating SiO$_2$ and Nb layers 0.7 $\mu$m tall. The top-most layer is Nb 0.2 $\mu$m thick. This pillar is part of the fill used to make the surface  flat on average to facilitate the chemical-mechanical polishing steps in producing our scanning SQUID susceptometers \cite{kirtley2016sss}. The image was taken at 4.2 K using a 1 mA amplitude, 500 Hz sinusoidally varying current through the field coil. The colormap scale on this image is labeled in units of the normalized change in susceptibility $\Delta\Phi/I\Phi_0$. The solid line in Fig.  \ref{fig:dot_susc_data}(b) displays a cross-section along the dashed line in Fig. \ref{fig:dot_susc_data}(a). The open circles are calculations following Sec. \ref{sec:calculation} with the spacing between the surfaces of the sample and susceptometer $z_{\rm fit}$ as a fitting parameter. The best fit was for $z_{\rm fit}=0.8$ $\mu$m. The calculation produces a positive overshoot to the susceptibility larger than is observed experimentally, and the spacing $z_{\rm exp}$ is larger than expected from the sample geometry and other results. We speculate that this discrepancy may be due to the 3-dimensional nature of the Nb/SiO$_2$ pillar. Nevertheless, the calculations reproduce the width of the susceptibility transition, about 1 $\mu$m from the 10\% to 90\% points.

\section*{Conclusions}

In conclusion, we have introduced a method for calculating the response of a deep sub-micron pickup loop scanning SQUID susceptometer to external fields by solving coupled London's and Maxwell's equations. The results of this calculation agree reasonably well with experiments using a SQUID susceptometer with a deep sub-micron sized pickup loop for various sources of magnetic field. These calculations should provide a useful tool for interpreting scanning SQUID microscopy data with sub-micron spatial resolution.

\section*{Acknowledgements}

This work was supported by an NSF IMR-MIP Grant No. DMR-0957616. J.C.P. was supported by a Gabilan Stanford Graduate Fellowship and an NSF Graduate Research Fellowship, Grant No. DGE-114747. The nanomagnet sample fabrication at Cornell was supported by the NSF (DMR-1406333 and through the Cornell Center for Materials Research, DMR-1120296), and made use of the Cornell Nanoscale Facility which is supported by the NSF (ECCS-1542081). We would like to thank Micah J. Stoutimore for providing the Nb film used to study vortices.

\section*{Citations}
\bibliographystyle{unsrt} 
\bibliography{pairing_symmetry}

\end{document}